\documentclass[showpacs,preprintnumbers,amsmath,amssymb]{revtex4}
\newtheorem{theorem}{Theorem}
\newtheorem{definition}{Definition}
\newtheorem{lemma}{Lemma}
\newtheorem{corollary}{Corollary}

\usepackage{graphicx}
\usepackage{dcolumn}
\usepackage{bm}

\begin{document}


\title{
Heisenberg's uncertainty principle
 for simultaneous measurement of positive-operator-valued measures}

\author{Takayuki Miyadera}
\email{miyadera-takayuki@aist.go.jp}
\author{Hideki Imai}%
 \altaffiliation[Also at ]{Graduate School of Science and Engineering,
Chuo University.
1-13-27 Kasuga, Bunkyo-ku, Tokyo 112-8551, Japan .
 }%
\affiliation{%
Research Center for Information Security (RCIS), \\
National Institute of Advanced Industrial
Science and Technology (AIST). \\
Daibiru building 1102,
Sotokanda, Chiyoda-ku, Tokyo, 101-0021, Japan.
}%


\date{\today}

\begin{abstract}
A limitation on simultaneous measurement of two 
arbitrary positive operator valued measures is discussed. 
In general, simultaneous measurement of two noncommutative 
observables is only approximately possible. 
Following Werner's formulation, we introduce a distance 
between observables to quantify an accuracy of measurement. 
We derive an inequality that relates 
the achievable accuracy with noncommutativity between 
two observables. 
As a byproduct a necessary condition for 
two positive operator valued measures to be simultaneously 
measurable is obtained.    
\end{abstract}

\pacs{03.65.Ta, 03.67.-a}
\maketitle

\section{Introduction}
Heisenberg's uncertainty principle is often considered as 
one of the most important features of quantum theory. 
In every text book on quantum theory one can find 
its explanation proposed by 
Heisenberg himself \cite{Heisenberg} 
and its ``derivation" by Robertson \cite{Robertson}.  
However, as recently claimed by several researchers, 
the above explanation and the derivation have 
a certain gap between them. On the one hand 
Heisenberg is concerned with a simultaneous measurement
of position and momentum, on the other hand the Robertson's
formulation treats two distinct measurements each on position and 
momentum. 
The later formulation has been actively investigated 
since then, and there are a several different inequalities
for general observables 
depending on quantities that characterize the 
uncertainty of probability distributions
\cite{LP,Deutsch,Maassen,KP,Miyadera}.  
In contrast to it, it seems quite recent that 
the former formulation began being
reflected extensively. In this formulation
one must deal with a simultaneous (or joint) measurement of two 
or more observables and somehow estimate its 
accuracy compared with 
individual measurements of the observables. 
Appleby, in his pioneering work \cite{Appleby}, 
investigating simultaneous measurement of 
position and momentum of quantum mechanical particles, 
introduced error operators 
and disturbance operators and derived 
simple inequalities between them. 
 Ozawa \cite{Ozawa} treated a pair of general 
 self-adjoint observables and 
 considered a tradeoff relation between his 
 error operator and disturbance operator that have 
 an interpretation in the context of his extended notion of 
 simultaneous measurement. 
 Werner \cite{Werner} formulated the problem from an operational viewpoint 
 and derived a beautiful inequality between 
 position and momentum operator of quantum mechanical particles.
 Janssens' work \cite{Janssens} is related with it and 
 showed a nice inequality on added variances 
 between arbitrary 
 self-adjoint observables in a simple manner.   
 Busch and Pearson \cite{errorbar} introduced a notion of 
 error bar that represents a resolution of the measurements, and 
 discuss a tradeoff relation related with position and 
 momentum. Busch and Heinosaari \cite{Heinosaari} estimated a tradeoff
 between observables in a single qubit in detail. 
 Busch, Heinonen and Lahti gave a review on this topic in \cite{BHL}.
  \par
  In this paper, we deal with simultaneous measurements 
  of two arbitrary observables (positive operator valued measures). 
 In general, a pair of noncommutative observables 
 is not simultaneously measurable. 
 We derive an inequality that relates the limitation on 
 the simultaneous measurements with the noncommutativity
 of observables. 
 The paper is organized as follow:
 In section \ref{sec:formulation}, we 
 explain our formulation of the problem. 
 For that purpose we  introduce
  notions of distance between 
 observables and simultaneous measurability of 
 observables. 
 In section \ref{sec:main}, we derive our main 
 result. Tradeoff relations related with 
 two types of distances are explained. 
 As a byproduct we give a necessary condition for 
 two positive operator valued measures to be 
 simultaneously measurable. 
 A simple example is also discussed. 
\section{Formulation}\label{sec:formulation}
In this paper, we follow a formulation introduced by Werner \cite{Werner}
which has a clear operational meaning. 
Suppose that we have a quantum system described by a Hilbert space 
${\cal H}$ and 
a pair of observables;
$A$ and $B$ on it. 
$A$ and $B$ are
represented as positive operator 
valued measures (POVMs) $A=\{A_a\}_{a \in \Omega_A}$ and 
$B=\{B_b\}_{b \in \Omega_B}$.
 That is,  
for each $a$ and $b$, $0\leq A_a, B_b\leq {\bf 1}$ holds and 
$\sum_a A_a =\sum_b B_b ={\bf 1}$ is satisfied. 
(For simplicity, we hereafter assume the sets of outcomes (say $\Omega_A$) 
finite.)
Given a state $\omega$, we can compute a probability distribution 
$\{\omega(A_a)\}_a$, where $P^{\omega}_{A}(a):=\omega(A_a)$
 is interpreted as a probability to 
obtain an outcome $a$ when we make a measurement on $A$ with 
respect to the state $\omega$. 
If $A_a$ and $B_b$ commute with each other for all $a$ and $b$, 
$A\circ B:=\{A_a B_b\}_{(a,b) \in \Omega_A \times \Omega_B}$ 
again defines a POVM. 
This newly defined POVM gives a probability distribution 
$P^{\omega}_{A\circ B}(a,b)=\omega(A_aB_b)$ whose marginal distributions 
coincide exactly with 
$P^{\omega}_{A}$ and $P^{\omega}_B$ for any state $\omega$. 
That is, in this case one can achieve a simultaneous measurement 
of $A$ and $B$ perfectly by using $A\circ B$. 
This is not always the case in general for noncommutative pairs. 
What we are interested in is that impossibility. 
What is the quantitative limitation on simultaneous measurements 
of general pairs of observables? 
\par
To formulate this problem quantitatively, one has to 
introduce a proper distance between two probability distributions. 
Suppose that we have a space ${\cal P}(\Omega)$ 
of probability distributions 
over a (finite) sample space $\Omega$. 
(i.e., ${\cal P}(\Omega):=\{p:\Omega \to [0,1]|\ \sum_x p(x)=1\}$)
\par
A function $d: {\cal P}(\Omega) \times {\cal P}(\Omega)
\to {\bf R}_{+}$ is called a distance if the following conditions are 
satisfied: 
\begin{itemize}
\item[(i)] $d(p,q)=d(q,p)$,
\item[(ii)]$d(p,p)=0$ $\Leftrightarrow$ $p\equiv 0$,
\item[(iii)] $d(q,p)+d(p,r)\geq d(q,r)$ (triangle inequality). 
\end{itemize}
For instance, 
$l_{\infty}$ distance (uniform distance) is defined as
\begin{eqnarray*}
d_{\infty}(p,q):=\max_{x\in \Omega} |p(x)-q(x)|,
\end{eqnarray*}
and $l_{1}$ distance is defined as 
\begin{eqnarray*}
d_1(p,q):=\frac{1}{2}
\sum_{x \in \Omega} |p(x)-q(x)|.
\end{eqnarray*}
 (For later convenience 
we put a coefficient $\frac{1}{2}$.)
Once we fix a distance $d$ over the probability spaces, 
we can define a distance between two observables. 
Suppose that POVMs $A=\{A_a\}_{a \in \Omega}$ and $B=\{B_b\}_{b \in \Omega}$ 
have an identical set of outcomes. We define a distance between 
$A$ and $B$ with respect to $d$ as
\begin{eqnarray*}
D_d(A,B):=\sup_{\omega} d(P^{\omega}_A,P^{\omega}_B),
\end{eqnarray*}
where the supremum is taken over all the states.
One can easily show that $D_d(A,B)=0$ if and only if $A=B$. 
All other conditions for distance also follow easily.
Thus $D_d$ becomes a distance between a pair of POVMs 
which have an identical set of outcomes. 
\par
Let us recall the previous example in which 
all the elements of $A=\{A_a\}_{a \in \Omega_A}$ 
and $B=\{B_b\}_{b \in \Omega_B}$ commute with each other. 
In such a case one can define $A\circ B$ 
that has $\Omega_A \times \Omega_B$ as the set of 
outcomes.  
If we define $f_A: \Omega_A \times \Omega_B \to \Omega_A$ as 
$f_A(a,b)=a$ and $f_B: \Omega_A \times \Omega_B \to \Omega_B$ as 
$f_B(a,b)=b$ for all $a\in \Omega_A$ and 
$b \in \Omega_B$, $A_a =\sum_{(a',b')}^{f(a',b')=a}A_{a'} B_{b'}$ 
and $B_b =\sum_{(a',b')}^{f(a',b')=b}A_{a'} B_{b'}$ hold. 
That is, truncating outcomes of $A\circ B$ with some functions 
reproduces both $A$ and $B$.  
This can be generalized to the following procedure.
Let us consider a POVM $F:=\{F_x\}_{x \in \Omega}$. 
$f_A: \Omega \to \Omega_A$ defines a new POVM 
$f_A(F):=\{f_A(F)_a\}_{a \in \Omega_A}$ as 
\begin{eqnarray*}
f_A(F)_a :=\sum_{x: f_A(x)=a} F_x.
\end{eqnarray*}
Also, an other function $f_B: \Omega
\to \Omega_B$ defines a POVM $f_B(F)$. 
Here it is appropriate to say that 
POVMs $f_A(F)$ and $f_B(F)$ are simultaneously 
measurable (through a POVM $F$). 
Actually, the following definition of the simultaneously 
measurable observables is a standard one (See e.g. \cite{BHL,Araki}).
\begin{definition}
A pair of POVMs $A=\{A_a\}_{a\in \Omega_A}$ and $B=\{B_b\}_{b \in \Omega_B}$
is called simultaneously measurable if and only if 
there exists a POVM $F=\{F_x\}_{x \in \Omega}$ and a pair of 
functions $f_A: \Omega \to \Omega_A$ and $f_B: \Omega \to \Omega_B$ 
such that $A=f_A(F)$ and $B=f_B(F)$ hold. 
\end{definition}
\par
Having introduced the notions of distance and simultaneous 
measurement, we can formulate our problem as follows.
For given arbitrary pair of POVMs $A=\{A_a\}$ and $B=\{B_b\}$, 
we choose a proper POVM $F=\{F_x\}$ and functions $f_A$ and $f_B$ 
that approximately
 reconstruct them. How close can we make
  $f_A(F)$ and $f_B(F)$ 
 to $A$ and $B$? 
 Once we fix $F$, $f_A$ and $f_B$, we can define 
 the closeness of them to $A$ and $B$ as $D_d(f_A(F),A)$ and 
 $D_d(f_B(F), B)$. 
 One may expect that if one chooses $F$ to make 
 $D_d(f_A(F),A)$ small, $D_d(f_B(F),B)$ becomes 
 large. This tradeoff relation is what we are interested in. 
 \section{Results}\label{sec:main}
 \subsection{Uncertainty principle in $l_{\infty}$ distance}
 To proceed with the analysis, we have to fix a distance $d$ first. 
 The simplest $l_{\infty}$ distance $d_{\infty}$ has a clear 
meaning and in addition makes 
 our problem tractable. Indeed, in this case the corresponding 
 distance between two observables $A=\{A_a\}_{a\in \Omega}$ 
 and $A'=\{A'_a\}_{a\in \Omega}$ 
 can be written as \cite{Heinosaari}
 \begin{eqnarray*}
 D_{d_{\infty}}(A,A')
 =\sup_{\omega} \max_{a} |P^{\omega}_A(a)-
 P^{\omega}_{A'}(a)|
 =\max_a \Vert A_a -A'_a\Vert,
 \end{eqnarray*}
 where $\Vert\cdot \Vert$ represents an operator norm. 
Suppose that we fix a POVM $F$ and $f_A$ and $f_B$. 
$D_{d_{\infty}}(A,f_A(F))
=\max_a \Vert A_a-f_A(F)_a\Vert$ and 
$D_{d_{\infty}}(B,f_B(F))
=\max_b \Vert B_b-f_B(F)_b\Vert$ naturally follow. 
\par
The second observation that makes our analysis easier is a 
representation of POVMs by completely positive maps (CP maps)
\cite{JanssensMaster}. 
When we have an Abelian von Neumann algebra ${\cal M}$ and a 
unital (completely \cite{completely})
 positive linear map $T:{\cal M} \to {\bf B}({\cal H})$, 
a decomposition of unity (POVM) 
in ${\cal M}$, ${\bf 1}=\sum_{x \in \Omega}m_x$ 
with $m_x \geq 0$, defines a POVM $\{T(m_x)\}_{x\in \Omega}$ in
${\bf B}({\cal H})$. 
The inverse is also true. 
When we have a POVM $F=\{F_x\}_{x\in \Omega}$, 
we define ${\cal M}_{\Omega}$ a set of all the diagonal 
matrices acting on ${\bf C}^{|\Omega|}$. 
For each $x\in \Omega$, 
a projection onto the canonical basis vector
$e_x:=|x\rangle \langle x|$ 
is an 
element of ${\cal M}_{\Omega}$ and $e:=\{e_x\}_{x\in \Omega}$ 
is a decomposition of unity in ${\cal M}_{\Omega}$.  
One can define $T:{\cal M}_{\Omega} \to {\bf B}({\cal H})$ 
as 
\begin{eqnarray*}
T(\sum_{x\in \Omega}c_x e_x)
=\sum_{x\in \Omega}c_x F_x.
\end{eqnarray*}
It is easy to see that thus defined $T$ is indeed a CP map. 
\par
Let us begin the analysis. 
Suppose we have a pair of POVMs $A=\{A_a\}_{a\in \Omega_A}$ 
and $B=\{B_b\}_{b \in \Omega_B}$. Take an 
arbitrary POVM $F=\{F_x\}_{x\in \Omega}$ and functions 
$f_A: \Omega \to \Omega_A$ and $f_B: \Omega \to \Omega_B$. 
We represent the POVM $F$ in terms of a CP map 
$T: {\cal M}_{\Omega} \to {\bf B}({\cal H})$. 
That is, $F_x=T(e_x)$ holds for each $x \in \Omega$.
Since a CP map is linear, if we define 
$
E^A_a:=\sum_{x}^{f_A(x)=a}e_x=f_A(e)
$ and 
$E^B_b:=\sum_{x}^{f_B(x)=b}e_x=f_B(e)$
for each $a\in \Omega_A$ and $b \in \Omega_B$, 
we have
\begin{eqnarray*}
f_A(F)_a&=& T(E^A_a) \\
f_B(F)_b&=& T(E^B_b).
\end{eqnarray*} 
Since what we are interested in is 
$D_{d_{\infty}}(f_A(F),A)$ and $D_{d_{\infty}}(f_B(F),B)$, 
we shall 
estimate $\Vert T(E^A_a)-A_a\Vert$ and $\Vert T(E^B_b)-B_b \Vert$. 
They are represented in simple forms
if one defines {\it error operators},
\begin{eqnarray*}
\epsilon^A_a&:=&T(E^A_a)-A_a \\
\epsilon^B_b&:=&T(E^B_b)-B_b. 
\end{eqnarray*}
That is, $D_{d_{\infty}}(f_A(F),A)=\max_a \Vert \epsilon^A_a\Vert$ 
and $D_{d_{\infty}}(f_B(F),B)=\max_b \Vert \epsilon^B_b\Vert$ hold. 
We consider a commutator 
$[T(E^A_a),T(E^B_b)]=[\epsilon^A_a+A_a,\epsilon^B_b+B_b]
=[\epsilon^A_a, \epsilon^B_b]
+[\epsilon^A_a, B_b]+[A_a,\epsilon^B_b]+[A_a,B_b]$.
Taking the norm on both sides of the equation, 
$[B_b,A_a]=[\epsilon^A_a,\epsilon^B_b]
+[\epsilon^A_a,B_b]+[A_a,\epsilon^B_b]+[T(E^B_b),T(E^A_a)]$, 
we have
\begin{eqnarray}
\Vert [B_b,A_a]\Vert 
&\leq & \Vert [\epsilon^A_a, \epsilon^B_b ]\Vert 
+\Vert [\epsilon^A, B_b]\Vert +\Vert [\epsilon^B_b, A_a]\Vert 
+\Vert[T(E^B_b), T(E^A_a)]\Vert, 
\label{eqnorm}
\end{eqnarray}
where we have used the triangle inequality for norm. 
Each term in the right hand side is bounded as follows. 
The first term is $\Vert[\epsilon^A_a, \epsilon^B_b]\Vert \leq 2
\Vert \epsilon^A_a\Vert \Vert \epsilon^B_b\Vert$.
The second term is, by use of a relation $\Vert X\Vert =\sup_{\psi;
\Vert \psi\Vert =1} |\langle \psi|X|\psi \rangle|$
for self-adjoint operator $X$, 
\begin{eqnarray*}
\Vert [\epsilon^A_a, B_b]\Vert &=&
\Vert i[\epsilon^A_a, B_b]\Vert \\
&=&\sup_{\psi}|\langle \psi|
[\epsilon^A_a, B_b]\psi\rangle|
\\
&\leq& 
2 \sup_{\psi} \langle \psi|(\epsilon^A_a)^2 |\psi \rangle^{1/2}
\langle \psi|(\Delta B_b)^2 |\psi\rangle^{1/2}
\\
&\leq&
 2 \Vert \epsilon^A_a\Vert \sup_{\psi}\langle 
\psi|(\Delta B_b)^2|\psi\rangle^{1/2},
\end{eqnarray*}
where $\Delta X:=X-\langle \psi|X|\psi \rangle$ and we used the 
Robertson uncertainty relation. 
Due to $0\leq B_b \leq {\bf 1}$, 
$\langle \psi|(\Delta B_b)^2 |\psi\rangle^{1/2}\leq 1/2$ holds and
we have
\begin{eqnarray*}
\Vert [\epsilon^A_a, B_b]\Vert\leq \Vert \epsilon^A_a\Vert. 
\end{eqnarray*} 
For the third term, we have in the similar manner, 
\begin{eqnarray*}
\Vert [\epsilon^B_b, A_a]\Vert \leq \Vert \epsilon^B_b\Vert.
\end{eqnarray*}
Thus (\ref{eqnorm}) is bounded as
\begin{eqnarray}
\Vert [A_a,B_b]\Vert
\leq 2\Vert \epsilon^A_a \Vert \Vert \epsilon^B_b \Vert 
+\Vert \epsilon^A_a \Vert +\Vert \epsilon^B_b \Vert 
+\Vert [T(E^A_a), T(E^B_b)]\Vert. 
\label{mouchoi}
\end{eqnarray}
To estimate the last term in the right hand side of the above 
inequality we use the following lemma proved by Janssens\cite{Janssens}.
\begin{lemma}
Let ${\cal A}$ and ${\cal B}$ be von Neumann algebras and 
$T: {\cal B} \to {\cal A}$ a CP map. 
Let $B, \tilde{B}$ be commuting Hermitian operators in ${\cal B}$, 
then, 
\begin{eqnarray*}
\Vert T(B^2)-T(B)^2\Vert^{1/2}
\Vert T(\tilde{B}^2)-T(\tilde{B})^2\Vert^{1/2}
\geq \frac{1}{2}\Vert [T(B),T(\tilde{B})]\Vert
\end{eqnarray*}
holds. 
\end{lemma}
The proof is done by a direct application of the 
Cauchy-Schwarz inequality for operators. 
Since our $E^A_a$ and $E^B_b$ commute with each other, 
we can apply the above lemma to our inequality to obtain
\begin{eqnarray}
\Vert [T(E^A_a),T(E^B_b)]\Vert 
\leq 2 \Vert T(E^A_a)-T(E^A_a)^2\Vert^{1/2} 
\Vert T(E^B_b)-T(E^B_b)^2\Vert^{1/2},
\label{Tcommutator}
\end{eqnarray}
where we utilized $(E^A_a)^2=E^A_a$ and $(E^B_b)^2 =E^B_b$. 
To delete $T(E^A_a)$ and $T(E^B_b)$ from the above inequality, we use
$T(E^A_a)=\epsilon^A_a +A_a$ and $T(E^B_b)=\epsilon^B_b +B_b$. 
It derives
\begin{eqnarray*}
\Vert T(E^A_a)-T(E^A_a)^2 \Vert 
&=&\Vert \epsilon^A_a +A_a -(\epsilon^A_a)^2 -A_a^2
-\epsilon^A_a A_a -A_a \epsilon^A_a
\Vert
\\
&\leq&
\Vert \epsilon^A_a -(\epsilon^A_a)^2 -\epsilon^A_a A_a -A_a \epsilon^A_a 
\Vert 
+ \Vert A_a -A_a^2\Vert,
\end{eqnarray*}
whose first term in the right hand side 
is further bounded as follows:
\begin{eqnarray*}
\Vert \epsilon^A_a -(\epsilon^A_a)^2 -\epsilon^A_a A_a -A_a \epsilon^A_a 
\Vert 
&=&
\Vert (1-A_a)\epsilon^A_a -\epsilon^A_a A_a -(\epsilon^A_a)^2\Vert
\\
&=&
\Vert [A_a, \epsilon^A_a] -A_a \epsilon^A_a + (1-A_a)\epsilon^A_a -
(\epsilon^A_a)^2\Vert 
\\
&\leq &
\Vert [A_a,\epsilon^A_a]\Vert 
+\Vert (1-2A_a -\epsilon^A_a) \epsilon^A_a\Vert.
\end{eqnarray*}
Here we use the relation $\Vert [A_a, \epsilon^A_a]\Vert 
\leq \Vert \epsilon^A_a\Vert$ and $1-2A_a -\epsilon^A_a
=1-(A_a +T(E^A_a))$ to derive
\begin{eqnarray*}
\Vert [\epsilon^A_a, A_a]\Vert 
+\Vert (1-2A_a -\epsilon^A_a) \epsilon^A_a\Vert
&\leq& \Vert \epsilon^A_a \Vert 
+\Vert 1-(A_a +T(E^A_a))\Vert \Vert \epsilon^A_a\Vert
\leq
2 \Vert \epsilon^A_a\Vert, 
\end{eqnarray*}
where we used a relation $\Vert 1- (A_a +T(E^A_a))\Vert \leq 1$
that is derived from $0\leq A_a +T(E^A_a)\leq 2 {\bf 1}$.
Finally we obtain
\begin{eqnarray}
\Vert T(E^A_a)-T(E^A_a)^2 \Vert
\leq 2\Vert \epsilon^A_a\Vert +\Vert A_a-A_a^2\Vert.
\label{TAbound}
\end{eqnarray}
We are ready to state the following theorem.
\begin{theorem}\label{maintheorem}
Suppose that we have a pair of POVMs $A=\{A_a\}_{a\in \Omega_A}$ 
and $B=\{B_b\}_{b\in \Omega_B}$. For any choice of a POVM 
$F=\{F_x\}_{x\in \Omega}$ and a pair of functions 
$f_A: \Omega \to \Omega_A$ and $f_B: \Omega \to \Omega_B$, 
\begin{eqnarray*}
&&2 D_{d_{\infty}}(A, f_A(F))D_{d_{\infty}}(B, f_B(F))
+D_{d_{\infty}}(A,f_A(F))+D_{d_{\infty}}(B,f_B(F))
\\
&&
+2 (2D_{d_{\infty}}(A,f_A(F))+ V(A))^{1/2}(2D_{d_{\infty}}(B,f_B(F))
+V(B))^{1/2}\geq \max_{a,b}\Vert[A_a,B_b]\Vert
\end{eqnarray*}
holds, where $V(A):=\max_a\Vert A_a-A_a^2\Vert$ represents 
an intrinsic uncertainty of a POVM $A$ (and similarly for $V(B):=
\max_b \Vert B_b -B_b^2\Vert$). 
\end{theorem}
{\bf Proof:}
Combining (\ref{mouchoi}), (\ref{Tcommutator}) and 
(\ref{TAbound}), we obtain
\begin{eqnarray}
\Vert [A_a,B_b]
\Vert &&
\leq 2\Vert \epsilon^A_a \Vert \Vert \epsilon^B_b \Vert 
+\Vert \epsilon^A_a \Vert +\Vert \epsilon^B_b \Vert 
\nonumber
\\
&&
+2(2\Vert \epsilon^A_a\Vert +\Vert A_a-A_a^2\Vert
)^{1/2}
(2\Vert \epsilon^B_b\Vert +\Vert B_b-B_b^2\Vert
)^{1/2}.
\label{proof}
\end{eqnarray}
 We take its maximum over  
$a$ and $b$ to obtain the theorem.
\hfill Q.E.D. 
\par
The intrinsic uncertainty of a POVM $A$ satisfies 
$0\leq V(A)\leq \frac{1}{4}$. 
As corollaries, we obtain some observations.
\begin{corollary}\label{forPVM}
Suppose we have a pair of projection valued measures (PVMs) 
$A=\{A_a\}_{a\in \Omega_A}$ and $B=\{B_b\}_{b \in \Omega_B}$. 
For any choice of a POVM 
$F=\{F_x\}_{x\in \Omega}$ and a pair of functions 
$f_A: \Omega \to \Omega_A$ and $f_B: \Omega \to \Omega_B$, 
\begin{eqnarray*}
&&2 D_{d_{\infty}}(A, f_A(F))D_{d_{\infty}}(B, f_B(F))
+D_{d_{\infty}}(A,f_A(F))+D_{d_{\infty}}(B,f_B(F))
\\
&&
+4 D_{d_{\infty}}(A,f_A(F))^{1/2}D_{d_{\infty}}(B,f_B(F))^{1/2}
\geq \max_{a,b}\Vert[A_a,B_b]\Vert
\end{eqnarray*}
holds.
\end{corollary}
{\bf Proof:}
For projection $P$, $P =P^2$ and thus $V(A)=V(B)=0$ hold.
\hfill Q.E.D.
\par
From this corollary one can see that 
a pair of PVMs that have noncommutative elements 
is not simultaneously measurable. 
This tradeoff relation is, what we may call, 
Heisenberg's uncertainty principle. 
\par
On the other hand, it is possible for a pair of POVMs 
to be simultaneously measurable even if they are 
noncommutative with each other. 
\begin{corollary}
Suppose that we have a pair of POVMs $A=\{A_a\}_{a\in \Omega_A}$ 
and $B=\{B_b\}_{b\in \Omega_B}$. 
If they are simultaneously measurable, their 
intrinsic uncertainties $V(A)$ and $V(B)$ satisfy
\begin{eqnarray*}
V(A)^{1/2}V(B)^{1/2}\geq \frac{1}{2} \max_{a,b}\Vert [A_a, B_b]\Vert.
\end{eqnarray*}
\end{corollary}
{\bf Proof:}
Put $D_{d_{\infty}}(f_A(F),A)=D_{d_{\infty}}(f_B(F),B)=0$ in 
Theorem \ref{maintheorem}.
\hfill Q.E.D.
\par
The simultaneous measurability of noncommutative POVMs 
is not surprising at all. In fact, suppose that we have 
a doubly indexed POVM $F=\{F_{(a,b)}\}_{(a,b)\in \Omega_A \times \Omega_B}$, 
we can construct a pair of 
simultaneously measurable
POVMs $f_A(F)$ and $f_B(F)$ 
by functions $f_A$ and $f_B$ with $f_A(a,b)=a$ and $f_B(a,b)=b$.
The above corollary says that any such a pair must have 
sufficiently large intrinsic uncertainties. 
\par
The following result is also easy to obtain.
We restrict the observable employed for the simultaneous 
measurement to PVM. 
\begin{corollary}
Suppose that we have a pair of POVMs $A=\{A_a\}_{a\in \Omega_A}$ 
and $B=\{B_b\}_{b\in \Omega_B}$. 
For any choice of a PVM
$F=\{F_x\}_{x\in \Omega}$ and a pair of functions 
$f_A: \Omega \to \Omega_A$ and $f_B: \Omega \to \Omega_B$, 
\begin{eqnarray*}
2 D_{d_{\infty}}(A, f_A(F))D_{d_{\infty}}(B, f_B(F))
+D_{d_{\infty}}(A,f_A(F))+D_{d_{\infty}}(B,f_B(F))
\geq \max_{a,b}\Vert[A_a,B_b]\Vert
\end{eqnarray*}
holds. 
\end{corollary}
{\bf Proof:}
In (\ref{mouchoi}), we can put $[T(E^A_a),T(E^B_b)]=0$.
\hfill Q.E.D.
\subsection{Uncertainty principle in $l_1$ distance}
Next we consider another distance $D_{d_1}$ induced by 
the $l_1$ distance. 
The following observation \cite{Werner} plays a crucial role 
for the analysis. For a pair of POVMs $A=\{A_a\}_{a\in \Omega}$
and $A'=\{A'_a\}_{a\in \Omega}$,  
\begin{eqnarray*}
D_{d_1}(A,A')=\sup_{\omega}\frac{1}{2}
\sum_{a\in \Omega}
|P^\omega_A(a)-P^\omega_{A'}(a)|
=\max_{\Delta \subset \Omega}
\Vert \sum_{a\in \Delta}A_a
-\sum_{a\in \Delta}A'_a\Vert
\end{eqnarray*}
holds. Thus if we define for each $\Delta_A \subset \Omega_A$ 
and $\Delta_B \subset \Omega_B$, 
$A_{\Delta_A}:=\sum_{a\in \Delta_A}A_a$, 
$E^A_{\Delta_A}:=\sum_{a\in \Delta_A}E^A_a$
 and 
$B_{\Delta_B}:=\sum_{b\in \Delta_B}B_b$, 
$E^B_{\Delta_B}:=\sum_{b\in \Delta_B}E^B_b$, 
error operators should be introduced as
$\epsilon^A_{\Delta_A}:=T(E^A_{\Delta_A})-A_{\Delta_A}$
and $\epsilon^B_{\Delta_B}:=T(E^B_{\Delta_B})-
B_{\Delta_B}$, and the analysis to obtain 
equation (\ref{proof})
works 
just by replacing as follows:
\begin{eqnarray*}
A_a &\to& A_{\Delta_A} \\
B_b &\to& B_{\Delta_B} \\
E^A_a &\to& E^A_{\Delta_A}\\
E^B_b &\to& E^B_{\Delta_B} \\
\epsilon^A_a &\to& \epsilon^A_{\Delta_A}\\
\epsilon^B_b &\to& \epsilon^B_{\Delta_B}.
\end{eqnarray*}
That is, it holds
\begin{eqnarray*}
\Vert [A_{\Delta_A},B_{\Delta_B}]
\Vert &&
\leq 2\Vert \epsilon^A_{\Delta_A}
 \Vert \Vert \epsilon^B_{\Delta_B} \Vert 
+\Vert \epsilon^A_{\Delta_A} \Vert +\Vert \epsilon^B_{\Delta_B}
 \Vert 
\\
&&
+2(2\Vert \epsilon^A_{\Delta_A}
\Vert +\Vert A_{\Delta_A}-A_{\Delta_A}^2\Vert
)^{1/2}
(2\Vert \epsilon^B_{\Delta_B}\Vert +\Vert B_{\Delta_B}
-B_{\Delta_B}^2\Vert
)^{1/2}.
\end{eqnarray*}
Taking the maximum over the subsets $\Delta_A$ and $\Delta_B$, 
we obtain the following theorem.
\begin{theorem}
Suppose that we have a pair of POVMs $A=\{A_a\}_{a\in \Omega_A}$ 
and $B=\{B_b\}_{b\in \Omega_B}$. For any choice of a POVM 
$F=\{F_x\}_{x\in \Omega}$ and a pair of functions 
$f_A: \Omega \to \Omega_A$ and $f_B: \Omega \to \Omega_B$, 
\begin{eqnarray*}
&&2 D_{d_{1}}(A, f_A(F))D_{d_{1}}(B, f_B(F))
+D_{d_{1}}(A,f_A(F))+D_{d_{1}}(B,f_B(F))
\\
&&
+2 (2D_{d_{1}}(A,f_A(F))+ V_1(A))^{1/2}(2D_{d_{1}}(B,f_B(F))
+V_1(B))^{1/2}
\\
&&
\geq \max_{\Delta_A \subset 
\Omega_A, \Delta_B\subset \Omega_B}\Vert
\sum_{a\in \Delta_A}
\sum_{b \in \Delta_B} [A_a,B_b]\Vert
\end{eqnarray*}
holds, where $V_1(A):=\max_{\Delta_A\subset 
\Omega_A}\Vert \sum_{a\in \Delta_A}
A_a ({\bf 1}-\sum_{a\in \Delta_A} A_a)\Vert$ represents 
an intrinsic uncertainty of a POVM $A$ (and similarly for $V_1(B)$). 
\end{theorem}
The corresponding corollaries can be derived easily.
For instance, the following statement hold.
\begin{corollary}
Suppose we have a pair of projection valued measures (PVMs) 
$A=\{A_a\}_{a\in \Omega_A}$ and $B=\{B_b\}_{b \in \Omega_B}$. 
For any choice of a POVM 
$F=\{F_x\}_{x\in \Omega}$ and a pair of functions 
$f_A: \Omega \to \Omega_A$ and $f_B: \Omega \to \Omega_B$, 
\begin{eqnarray*}
&&2 D_{d_{1}}(A, f_A(F))D_{d_{1}}(B, f_B(F))
+D_{d_{1}}(A,f_A(F))+D_{d_{1}}(B,f_B(F))
\\
&&
+4 D_{d_{1}}(A,f_A(F))^{1/2}D_{d_{1}}(B,f_B(F))^{1/2}
\geq 
\max_{\Delta_A \subset 
\Omega_A, \Delta_B\subset \Omega_B}\Vert
\sum_{a\in \Delta_A}
\sum_{b \in \Delta_B} [A_a,B_b]\Vert
\end{eqnarray*}
holds.
\end{corollary}
%
\subsection{Example: A qubit}
\begin{figure}[thbp]
\begin{center}
 \includegraphics[width=40mm]{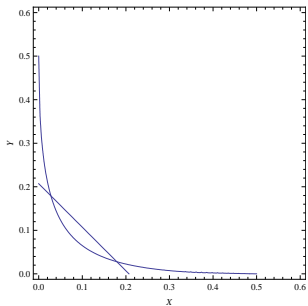}
\end{center}
 \caption{ \label{fig:one}
 Admissible region of $X:=D_{d_0}(f_A(F),A)$ ($x$-axis) and 
 $Y:=D_{d_0}(f_B(F),B)$ ($y$-axis) for $\theta=\frac{\pi}{2}$. 
 A curved line is a contour of admissible region 
 obtained by our method. A straight line is a contour of 
 admissible region obtained in (\ref{eqHeino}).
 }
\end{figure}

As the most simple example, we study a pair of 
PVMs for a single qubit. 
Each projection operator is parameterized by a Bloch sphere as, 
$E({\bf n}):=\frac{1}{2}({\bf 1}+{\bf n}\cdot {\bf \sigma})$
for ${\bf n}\in {\bf R}^3$ with $|{\bf n}|=1$. 
Let us consider two PVMs, 
$A_{\bf n}=\{E({\bf n}),E(-{\bf n})\}$ and 
$A_{\bf m}=\{E({\bf m}), E(-{\bf m})\}$ with 
$\angle({\bf n},{\bf m})=\theta \ (0\leq \theta \leq \frac{\pi}{2})$. 
They satisfy the following inequality. 
\begin{corollary}
For any POVM $F$ and functions $f_{A_{\bf n}}$ and $f_{A_{\bf m}}$, 
\begin{eqnarray*}
&&2 D_{d_{\infty}}(A_{\bf n}, f_{A_{\bf n}}(F))
D_{d_{\infty}}(A_{\bf m}, f_{A_{\bf m}}(F))
+D_{d_{\infty}}(A_{\bf n},f_{A_{\bf n}}(F))
+D_{d_{\infty}}(A_{\bf m},f_{A_{\bf m}}(F))
\\
&&
+4 D_{d_{\infty}}(A_{\bf n},f_{A_{\bf n}}(F))^{1/2}
D_{d_{\infty}}(A_{\bf m},f_{A_{\bf m}}(F))^{1/2}
\geq \frac{\sin \theta}{2}
\end{eqnarray*}
holds. 
\end{corollary}
{\bf Proof:}
With $[E({\bf n}), E({\bf m})]=\frac{i}{2}{\bf \sigma}\cdot 
({\bf n} \times {\bf m})$, it is immediate from Corollary \ref{forPVM}. 
\hfill Q.E.D.
\\
This bound should compared with a bound obtained in 
\cite{Heinosaari}. 
They derived an inequality,
\begin{eqnarray}
D_{d_{\infty}}(A_{\bf n},f_{A_{\bf n}}(F))
+D_{d_{\infty}}(A_{\bf m},f_{A_{\bf m}}(F))
\geq \sqrt{\frac{1}{2}}
\left(\cos \frac{\theta}{2}+\sin \frac{\theta}{2}
-1
\right). 
\label{eqHeino}
\end{eqnarray}
In Figure \ref{fig:one} the contours 
of their admissible regions 
are shown. One can see that in some region ours is 
better and in other region worse. It may be interesting 
that our method gives a nonlinear estimate in contrast with 
\cite{Heinosaari}.
\section{Summary}
In this paper, 
a limitation on simultaneous 
measurement of two arbitrary (discrete) 
positive operator valued measures
was investigated. 
Following Werner's work \cite{Werner}, 
we introduced the distance between observables
 by using 
distance between probability distributions. 
Introduction of the error operators simplified
the analysis.
We derived a novel inequality (Theorem \ref{maintheorem}), a possible 
representation of Heisenberg's uncertainty 
principle, that relates the limitation with 
noncommutativity. 
As a byproduct, we obtained a 
corollary indicating a necessary condition for 
a pair of POVMs to be simultaneously measurable.
Compared with the previous works on this subject, 
the broad applicability of our result to an arbitrary
discrete pair of POVMs is an advantage.
Extension of our result to other distances will be a future problem. 
\\
{\bf Acknowledgments}
\par
The authors thank an anonymous referee for fruitful comments. 


\begin{thebibliography}{9}
\bibitem{Heisenberg}W.~Heisenberg, Z. Phys. {\bf 43}, 172 (1927).
\bibitem{Robertson}H.~P.~Robertson, Phys. Rev. {\bf 34}, 163 (1929).
\bibitem{LP}H.~J.~Landau and H.~O.~Pollak, 
Bell Syst. Tech. J. {\bf 40}, 65 (1961).
\bibitem{Deutsch}D.~Deutsch, 
Phys. Rev. Lett. {\bf 50}, 631 (1983).
\bibitem{Maassen}H.~Maassen and J.~B.~M.~Uffink, 
Phys. Rev. Lett. {\bf 60}, 1103 (1988).
\bibitem{KP}M.~Krishna and K.~R.~Parthasarathy, 
Sankya, Ser. A {\bf 64}, 842 (2002).
\bibitem{Miyadera}T.~Miyadera and H.~Imai, 
Phys. Rev. A {\bf 76}, 062108 (2007).
\bibitem{Appleby}
D.~M.~Appleby,
Int.J.Theor.Phys. {\bf 37}, 1491 (1998); 
{\em ibid}. {\bf 37} 2557 (1998).
\bibitem{Ozawa}M.~Ozawa, 
Ann. Phys. {\bf 311}, 350 (2004). 
\bibitem{Werner}
R.~F.~Werner, Quantum Inf. Comp. {\bf 4}, 546 (2004).
\bibitem{Janssens}B.~Janssens, quant-ph/0606093. 
\bibitem{errorbar}
P.~Busch and D.~B.~Pearson, J. Math. Phys. {\bf 48},
082103 (2007).
\bibitem{Heinosaari}P.~Busch and T.~Heinosaari, 
Quantum Inf. Comp. {\bf 8}, 0797 (2008).
\bibitem{BHL}P.~Busch, T.~Heinonen and P.~Lahti, 
Phys. Rep. {\bf 452}, 155 (2007). 

\bibitem{Araki}H.~Araki, {\it Mathematical Theory of Quantum 
Fields}, (Oxford Press, 2000).
\bibitem{JanssensMaster}B.~Janssens, Master thesis, quant-ph/0503009.
\bibitem{completely}
A map $\Gamma$ from a von Neumann algebra 
${\cal A}$ to a von Neumann algebra ${\cal B}$ is called 
a unital completely 
positive map if (i) it is linear, (ii)
$\Gamma({\bf 1})={\bf 1}$ and
(iii) its extension 
$\Gamma \otimes \mbox{id}: 
{\cal A} \otimes M({\bf C}^N) 
\to {\cal B} \otimes M({\bf C}^N)$ 
($M({\bf C}^N)$ is a set of all the $N\times N$ matrices)
for any positive integer $N$, is positive. 
The condition (iii) is called complete positivity. 
If ${\cal A}$ 
is Abelian, the 
complete positivity of the map $\Gamma$ automatically 
follows from its positivity. 
\end{thebibliography}
\end{document}